# Multi-network Topology Underlying Individual Language Learning Success

Peilun Song[1,2], Shuguang Yang[3], Xiujuan Geng[1,2], Zhenzhong Gan[4], Suiping Wang[4], Gangyi Feng[1,2*]

1. Department of Linguistics and Modern Languages, The Chinese University of Hong Kong
2. Brain and Mind Institute, The Chinese University of Hong Kong
3. School of Psychology, South China Normal University
4. Philosophy and Social Science Laboratory of Reading and Development in Children and Adolescents (South China Normal University), Ministry of Education, China

**Funding**

The work described in this paper was supported by grants from the Research Grants Council of the Hong Kong Special Administrative Region, China (Project No.: 14614221, 14612923, 14621424, and C4001-23YF to Gangyi Feng), the National Natural Science Foundation of China (Project No. 32322090 to Gangyi Feng), and Key Research and Development Program of Guangdong, China (grant number: 2023B0303010004 to Suiping Wang).

**\* Corresponding author**
Gangyi Feng
Brain and Mind Institute
Department of Linguistics and Modern Languages
The Chinese University of Hong Kong
Shatin, N.T., Hong Kong SAR, China
g.feng@cuhk.edu.hk

**Abstract**

Adult language learning varies greatly among individuals. Traditionally associated with frontotemporal language regions, this variability is increasingly seen as stemming from distributed brain networks. However, the role of these networks and their topological organization in explaining these differences remains unclear. We hypothesize that graph-theory-based network analysis of intrinsic multimodal connectivities across multiple networks explains overall and component-specific variations in language learning. We tested this in 101 healthy adults who underwent resting-state fMRI, structural MRI, and diffusion tensor imaging before seven days of six artificial language training tasks. We identified one dominant general learning component shared across tasks and five task-specific ones. Cross-validated predictive models used multimodal multi-network graph-theoretic metrics to predict final learning outcomes (LO) and rates (LR). We significantly predicted the LO and LR of the general component, which were primarily contributed by dorsal attention and frontoparietal networks. Nodal local efficiency was the most consistent predictor, with additional contributions from node clustering coefficient and network centrality for LR, highlighting local robustness, mesoscale network segregation, and global influence in explaining individual differences. Only task-specific word learning LO was predictable, relying on default mode and frontoparietal hubs with high betweenness centrality and efficiency. These findings demonstrate that intrinsic network topologies underlie differences in language learning success, supporting a multiple-systems hypothesis in which attentional-control networks interact with default and subcortical systems to shape learning trajectories. This advances mechanistic understanding and paves the way for personalized language education.

# Introduction

Success in learning a new language is challenging and varies significantly among individuals, particularly in adulthood (Dörnyei and Skehan, 2003; Lightbown and Spada, 2021; Skehan, 1991). Some learners acquire a language relatively rapidly, while others struggle even after extensive exposure and training (Kidd et al., 2018). This variation occurs not only within each linguistic component but also across different process components, such as speech learning for non-native phonetic contrasts (e.g., /l/ and /r/) (Chandrasekaran et al., 2012; Feng et al., 2019, 2021b; Golestani and Zatorre, 2009; Guion and Pederson, 2007), lexical-semantic mapping and word learning (Davis and Gaskell, 2009; Gu and Johnson, 1996; Nation, 2001; Wang et al., 2003), and morphosyntax and grammar learning (Feng et al., 2021b; Friederici, 2011; Morgan-Short et al., 2012). This variability is not well-explained by exposure alone; instead, it points to underlying neurocognitive organizational characteristics that differ across individuals (Zatorre, 2013; Zatorre et al., 2012). Therefore, understanding these neural organizations and their interactions is a crucial scientific and practical objective.

Behavioral differences in learning those language components are not only related to language regions but also connected to multiple brain networks and their associated cognitive and learning abilities, such as working memory (Linck et al., 2014; Verhagen and Leseman, 2016; Wen and Li, 2019), attentional capacity (Issa and Morgan-Short, 2019), executive function and cognitive control (Miyake et al., 2000; Reetzke et al., 2016), and auditory processing ability (Antoniou et al., 2015; Banai and Ahissar, 2013; Saito et al., 2021). This convergence suggests that language learning draws on multiple interacting cognitive and memory systems rather than a single "language module" (Ashby and Maddox, 2011; Ashby and O'Brien, 2005; Minda et al., 2024; Squire, 2004, 1987). A primary goal of cognitive neuroscience of language learning is to investigate how individual-specific neural organization accounts for behavioral differences in learning trajectories and ultimate success within and across linguistic components.

Network neuroscience offers a principled framework for addressing this question by characterizing large-scale brain organization in terms of distributed network systems and their interconnections (Bassett and Sporns, 2017; Bullmore and Sporns, 2009; Sporns, 2013). Moving beyond region-centric accounts focused on canonical frontotemporal language areas, recent findings show that language learning and processing in adulthood recruit multiple networks, including the frontal-parietal control network (FPN), dorsal attention network (DAN), default mode network (DMN), medial temporal systems, subcortical basal ganglia circuits and cerebellum (Bradley et al., 2013; Chein and Schneider, 2005; Fedorenko and Thompson-Schill, 2014; Feng et al., 2021b; Gracia-Tabuenca et al., 2024; Kuhl, 2010). For example, the FPN supports domain-general control and adaptive task-set regulation, which facilitate hypothesis testing and rule abstraction (Cole et al., 2013; Duncan, 2010; Novick et al., 2005), also closely tied to rule-based speech category and grammar learning. The DAN supports selective attention, also required for speech cue weighting in novel phonetic category learning (Chandrasekaran et al., 2012; Corbetta and Shulman, 2002). The hippocampal-medial temporal structures support rapid encoding of exemplars and also facilitate forming novel word-meaning associations (Davis and Gaskell, 2009). These findings motivate a systems-level view in which attentional, control, memory, and language-related systems dynamically interact to support learning different language components.

While group-level neuroimaging studies have characterized which networks are engaged on average, they often overlook the extent of individual differences in learning success (Kievit et al., 2017; Yarkoni and Westfall, 2017). A growing literature indicates that neuroimaging-derived metrics, such as functional connectivity at rest (Finn et al., 2015; Tavor et al., 2016), task-evoked activation and connectivity (Greene et al., 2018), structural connectivity (Forkel et al., 2014; Saygin et al., 2016), and cortical morphology or myeloarchitecture (Fjell et al., 2015; Vandewouw et al., 2021), can prospectively explain individual differences in cognitive ability, including in language domains (Feng et al., 2021b; Kuhl, 2010; Meng et al., 2022; Morgan-Short et al., 2012). Predictive modeling approaches such as connectome-based predictive modeling (CPM) and multivariate machine-learning approaches have further uncovered that individual differences in connectivity patterns can forecast behavioral performance across domains, from attention and working memory to language and academic achievement (Feng et al., 2021b; Greene et al., 2018; Meng et al., 2022; Rosenberg et al., 2016; Shen et al., 2017). These developments highlight the value of individualized, brain-based models that move from group-level observation to individual outcome prediction, an essential step for mechanistic theory and translational application.

In this study, we combined network neuroscience approaches with predictive modeling methods to examine the extent to which large-scale neural network organization could explain individual differences in language learning success. The graph theory framework provides a quantitative method for characterizing how networks interact and communicate by analyzing brain network topologies (Bullmore and Sporns, 2009; Rubinov and Sporns, 2010), which enables us to link individual brain topological features to their success in language learning. Measures of network segregation (e.g., clustering coefficient, modularity), integration (e.g., global efficiency, characteristic path length), nodal influence and hubness (e.g., degree, betweenness centrality, PageRank), and local robustness and efficiency (e.g., local efficiency) capture complementary properties of information processing and inter-network communication (Betzel and Bassett, 2017; Sporns, 2013). Individual differences in these metrics have been associated with variation in executive function, memory, learning, and skill acquisition (Bassett et al., 2015; Cao et al., 2014; Li et al., 2009; Litwińczuk et al., 2023; Liu et al., 2025; Moreira Da Silva et al., 2020; Wang et al., 2016). In the language domain, brain connectivity within and beyond classical language areas has been related to phonological processing, vocabulary growth, and second-language proficiency (Chai et al., 2016; Feng et al., 2021b; Wei et al., 2012). These findings imply that individual differences in large-scale brain network organization, encompassing multiple networks and diverse structural, functional, and morphological features, may collectively explain the variability in language learning success.

A significant methodological and conceptual challenge in understanding individual differences in language learning is its multifaceted nature. Learning tasks that target phonetic category, lexical-semantic, and morphosyntax learning share some common cognitive demands (e.g., attention, working memory) and learning mechanisms, but also involve partly distinct processes and neural resources (Chandrasekaran et al., 2014; Davis and Gaskell, 2009; Ullman, 2004). This raises the risk of conflating linguistic component-specific variance with the variance related to overall language learning ability. Using methods like principal component analysis (PCA) enables us to estimate a data-driven latent component that captures shared variance across language tasks, as well as variance specific to each task, and provides several benefits. First, a general factor can index a stable, trait-like aptitude for language learning that is not tied to a particular stimulus set or training paradigm (Carroll, 1993; Conway and Kovacs, 2013; Kievit et al., 2017). Second, focusing prediction on the shared component reduces measurement error and task idiosyncrasies, increasing reliability and out-of-sample generalizability (Hedge et al., 2018; Spearman, 1904; Yarkoni and Westfall, 2017). Third, it aligns with multiple-systems accounts by allowing us to test whether network-level neuromarkers map onto a component-general language learning capacity that supports phonological, lexical, and grammatical learning (Davis and Gaskell, 2009; Dehaene-Lambertz et al., 2006; Ullman, 2004). Finally, comparing task-general and task-specific language learning components allows a separation of network contributions, such as whether the FPN and DAN mainly reflect general language learning ability, while frontotemporal language regions are linked to component-specific variations.

Here, we investigated the extent to which multiple networks and their multimodal brain topology predict individual differences in language learning at both component-general and component-specific levels. Before an artificial language training, 101 healthy adults underwent structural MRI (sMRI), diffusion tensor imaging (DTI), and resting-state functional MRI (rs-fMRI). Participants then completed seven days of training on six language learning tasks, encompassing auditory and speech category learning, word learning, morphosyntax, and learning phrase and sentence structure (see Fig. 1 for task descriptions). We constructed morphological, structural, and functional connectivity matrices from the three imaging modalities and computed a suite of graph-theoretic metrics that quantify complementary aspects of network topology ranging from local to global network topological features (Fig. 2A). We analyzed these multi-network topologies across cortical and subcortical networks at both group and individual levels, and further described the network variabilities among learners. Using multivariate predictive modeling and a cross-validation procedure, we tested whether these network topologies could predict component-general and specific learning performances that reflect differences among learners. We focused on individual learners' final learning outcomes and learning rates. By integrating multimodal connectomes with the component decomposition of individual learning curves, we identified topological neuromarkers from multiple networks, beyond a single network, that predict success across language learning components. These findings potentially advance mechanistic understanding and inform personalized strategies for language education and rehabilitation.

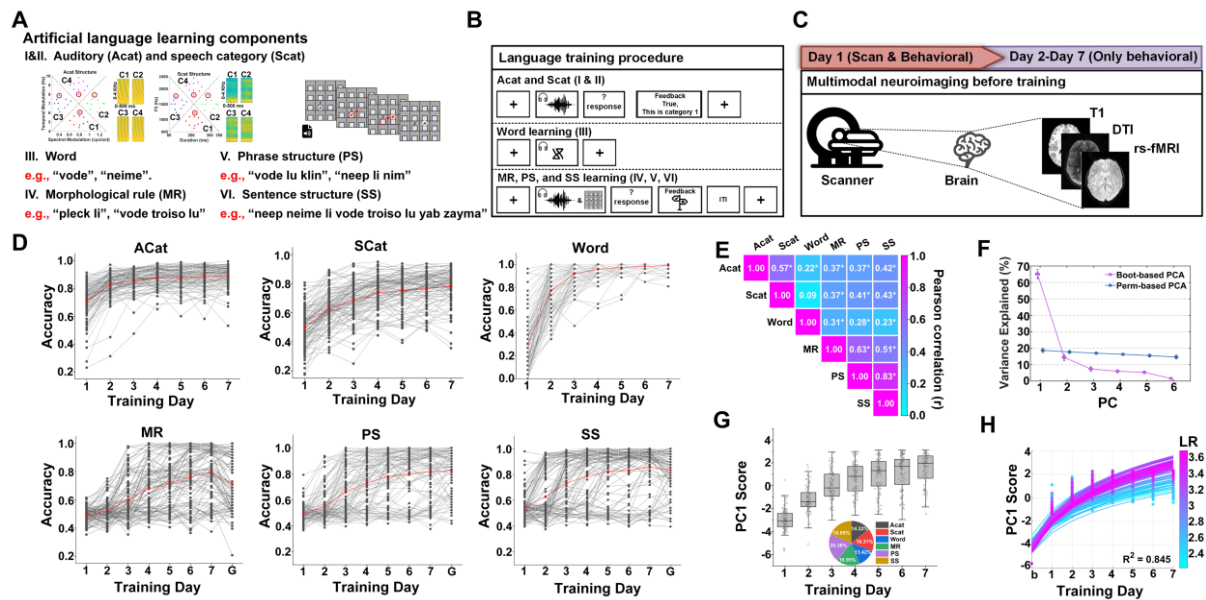

**Fig. 1**. Artificial language training materials, procedures, behavioral performance, and decomposition of learning curves across six tasks. **A**. Artificial language training stimuli for learning words, morphological rules (MR), phrase structures (PS), and sentence structures (SS), along with auditory (Acat) and speech (Scat) category structures. For both Acat and Scat, four categories (C1-C4) and example spectrograms were displayed in an acoustic space, with the optimal boundaries (dashed lines) shown. In the illustrated trial in SS, the spoken sentence meaning, "The square neep horizontally releases the round vode," is paired with corresponding visual movements on the chessboard. **B**. Learning procedures for the six tasks, where the two category-training tasks used the same procedure and the three grammar tasks follow the same procedure. **C**. Multimodal neuroimaging sessions before training. T1, DTI, and rs-fMRI data were acquired for each learner on Day 1, prior to training. **D**. Day-by-day learning performances for the six tasks (Acat, Scat, Word, MR, PS, and SS). Thin gray lines = individuals; red lines = group means; G = generalization test on Day 7. **E**. Correlation matrix of final-day learning performances between the six tasks. *, FDR corrected $p < 0.05$. **F**. Variance explained by each PCA component. PC1 exceeds the permutation-based null distribution ($p < 0.05$). **G**. PC1 scores across training days (boxplots) and tasks (pie chart). The pie chart shows task contributions to PC1, indicating nearly uniform loadings across tasks. **H**. Individual PC1 learning curves were fitted with a linear-log function. The models closely matched the learning curves (mean $R^2 = 0.845$); the colors indicate the estimated learning rates, with b representing baseline chance performance.

## Results

### Language learning performance across tasks

A total of 101 adults were included in the analyses. They completed six learning tasks over seven training days (see procedures in Materials and Methods; schematic in Fig. 1A-C). Learning performances were evaluated with: (i) two category-learning tasks focusing on auditory ripple categories (Acat) and speech vowel categories (Scat), (ii) a vocabulary training and recall test (Word), and (iii) three grammaticality judgment training tasks with feedback, learning morphological rules (MR; i.e., gender marker), Phrase structure (PS; i.e., subject-verb order), and sentence structure (SS; i.e., subject-object-verb order). All tasks showed consistent improvement across days, with notable variability between individuals within and across tasks (Fig. 1D). Moderate correlations were found between the final performances on the six training tasks (Fig. 1E), with stronger correlations within the same type (e.g., the two category learning tasks) than across different types (e.g., between category and word learning tasks).

To characterize shared and task-specific variance, we used principal component analysis (PCA) on the 7-day performance matrix (participants × days × tasks). The first principal component (PC1) explained significantly more variance (65% of the variance explained) than expected under permutation ($p < 0.0001$; Fig. 1F). PC1 scores increased monotonically over the training period (Fig. 1G), indicating a general language learning factor that reflected participants' overall progress in learning. Task loadings on PC1 were similar in magnitude across all six tasks (pie chart in Fig. 1G), indicating a cognitive construct of a broad, task-general learning factor. We also identified other components (PCs 2-6) that captured task-specific variances, ranging from 2% to 14% (see Supplementary Fig. S2 for the relative contribution of each task to these PCs).

To quantify individual learning rates, we modeled each learner's PC1 trajectories using a linear-log growth function. The models closely matched the observed PC1 learning curves (mean $R^2 = 0.845$), capturing both the early rapid improvements and the subsequent leveling off from baseline (chance) to Day 7 (Fig. 1H). The parameters from the models were used to determine each learner's learning rate (LR). The Day 7 PC1 scores served as the individual learning outcomes (LO). We defined the LR and LO the same way for other PCs.

### Multimodal brain network topology organization at the individual and group levels

We derived nine graph-theoretic brain topological metrics from three modality-specific connectivity matrices, including morphological (T1 radiomics similarity), structural (DTI FA-weighted tractography), and resting-state functional (rs-fMRI time-series correlation) matrices. These metrics characterized the topological organization of brain networks, ranging from local (e.g., degree centrality [DC]) to global (e.g., betweenness centrality [BC]) scales (Fig. 2A and B). Across the three neuroimaging modalities, the nine metrics exhibited wide variation between individuals, indicating significant individual differences in network topological structure (see Fig. 2C for the network topological profiles of sample subjects).

At the group level, surface maps and network summaries revealed distinct spatial patterns of the metrics across imaging modalities (see Fig. 2D for two representative metrics). Degree-based local topological metrics showed higher values in primary sensorimotor and visual regions for DTI, aligning with densely myelinated pathways. In contrast, rs-fMRI highlighted hubs within control and attention networks. The global topological metrics, such as betweenness centrality (BC), were more prominent in association networks compared to sensorimotor networks, with modality-specific peaks, such as frontoparietal and dorsal attention networks being most notable in rs-fMRI, while default mode and limbic areas contributed more in T1-derived morphology. These patterns emphasize that network topology depends jointly on imaging modality and brain network, which indicates the complementary roles across modalities in characterizing individual- and group-level brain topological structures.

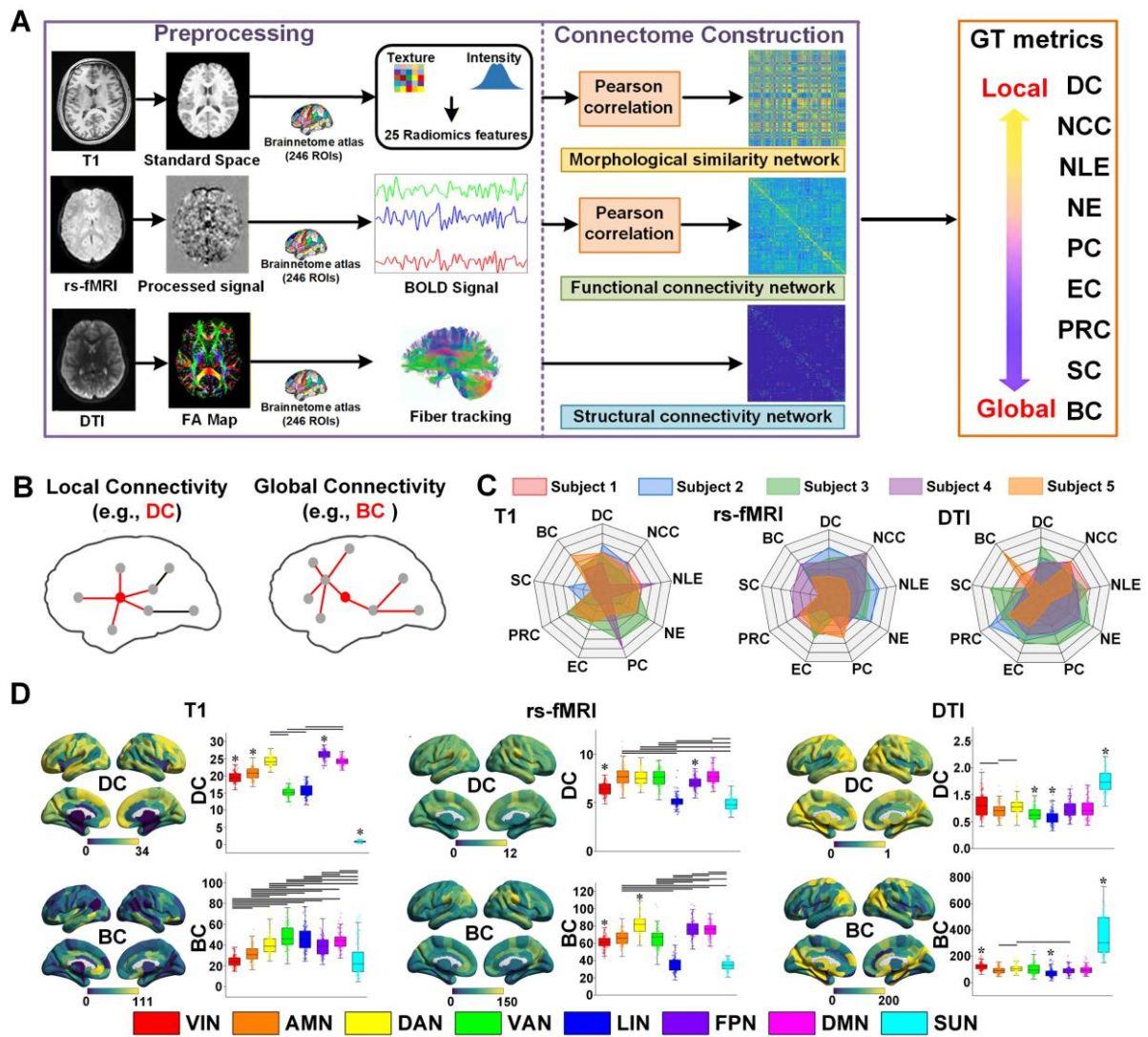

**Fig. 2**. Construction procedure of multimodal connectivity matrix and the imaging modality-dependent organization of brain network topology. **A**. Brain topological analysis pipeline for each imaging modality. T1 radiomics features, rs-fMRI regional time series, and DTI tractography were utilized to construct morphological, functional, and structural connectivity matrices, respectively. Nine graph-theory metrics were computed for each modality. **B**. Metrics can be categorized as those focusing on local (e.g., degree centrality [DC]) and those emphasizing global (e.g., betweenness centrality [BC]) connectivity. **C**. Individual differences in brain topology profiles. Radar plots for five representative learners show normalized values for all nine metrics averaged across modalities, highlighting significant variation between individuals in brain topological patterns. **D**. Spatial distributions and network-wide comparisons for BC and DC are shown. Asterisks indicate networks significantly different from all others; horizontal bars denote significant pairwise differences (see Supplementary Fig. S3 for other metrics). Network abbreviations: VIN = visual network; AMN = auditory-motor

network; DAN = dorsal attention network; VAN = ventral attention network; LIN = limbic network; FPN = frontal-parietal network; DMN = default mode network; SUN = subcortical network.

**Patterns of brain network topology across modalities**

We used PCA to reveal general patterns of brain network topology across the three imaging modalities and nine graph-theoretic metrics (data matrix: 101 participants × 264 regions × 3 modalities × 9 metrics). Two components captured a significant amount of variance (70.7% in total), with PC1 explaining 49.8% and PC2 explaining 20.9% of the total variance (both PCs better than chance with $p < 0.0001$; Fig. 3A).

To understand the two dimensions, we examined the composition of the two PCs at the graph network metric level. Loadings in the PC1-PC2 space showed that local network segregation metrics mainly contributed to PC1, such as node local efficiency (NLE), node clustering coefficient (NCC), and degree centrality (DC). PC2 was primarily influenced by global network integration metrics, including betweenness centrality (BC), PageRank centrality (PRC), and eigenvector centrality (EC) (Fig. 3B). Consequently, PC1 indicates regional segregation or local connectivity, whereas PC2 reflects global integration or communication between networks. We also found a negative correlation between the two PCs ($r = -0.68$, $p = 0.04$), suggesting this dissociation pattern.

Projecting the eight functional brain networks into the PC1-PC2 space further revealed a cortico-subcortical division, with cortical networks such as the frontal-parietal network (FPN), dorsal attention network (DAN), and default mode network (DMN) clustering along the PC1 axis. In contrast, the subcortical network (SUN) aligned with PC2 (Fig. 3C). Cortical surface maps and network summaries further revealed distinct spatial profiles (Fig. 3E-F) between the two PCs. PC1 showed higher scores in cortical association systems, with peaks in the FPN, DAN, and DMN networks, and lower scores in subcortical regions. PC2 displayed the opposite pattern, with its strongest positive contribution in the subcortical network (SUN) and relatively lower scores in the association networks. Network comparisons (Fig. 3F) in PC loading confirmed significant differences among networks for each PC, with association cortical networks ranking highest on PC1 and SUN highest on PC2 (all pairwise tests FDR-corrected, $p < 0.05$).

Projecting individual learners in the component space revealed significant inter-learner variability and a negative relationship (Fig. 3D). It suggested a potential topological fingerprint of individual learners, highlighting that individual learners vary in the balance between network segregation (PC1) and integration (PC2) as well as inter-network communications between cortical (PC1) and subcortical (PC2) regions.

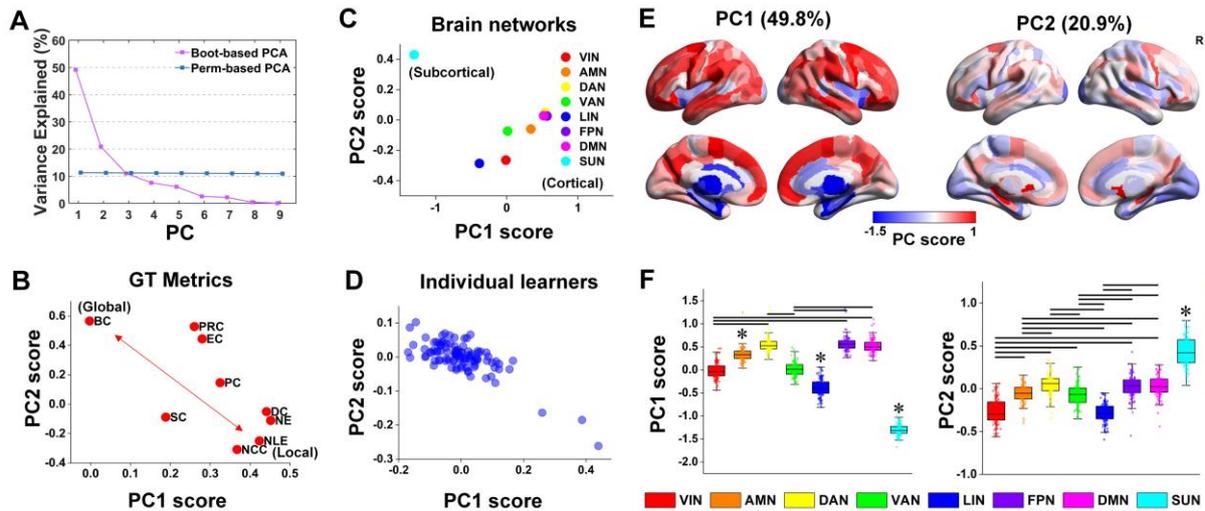

**Fig. 3**. Principal components of multimodal network topology reveal segregation-integration network organizations and cortical-subcortical network distinctions. **A**. PCA on nine graph metrics across three imaging modalities identified two significant principal components (PC1 = 49.8%, PC2 = 20.9% of variance; permutation test $p < 0.0001$). **B**. Loadings of the nine metrics in the PC1-PC2 space: PC1 is dominated by local-connectivity metrics (e.g., NLE, NCC, DC); PC2 by global-integration metrics (e.g., BC, PRC, EC). **C**. Projection of brain networks into the PC space shows association networks clustering along PC1, with the subcortical network aligning along PC2. **D**. Distribution of the 101 learners in the PC space highlights marked inter-individual variability along both cortical segregation (PC1) and subcortical integration (PC2) dimensions. **E**. Surface maps show PC1 (left) concentrated in cortical association regions and PC2 (right) emphasizing subcortical regions. R = right hemisphere. **F**. Network-wise comparisons: FPN, DAN, and DMN score highest on PC1, whereas subcortical networks (SUN) score highest on PC2; asterisks indicate networks differing from all others; horizontal bars mark significant pairwise contrasts (FDR $p < 0.05$).

**Language-component-general learning predictions**

We used a nested 5-fold cross-validation (CV) approach to construct and validate learning prediction models (Fig. 4A), examining the predictive ability of multimodal brain topology metrics on language learning success (LO and LR). In the prediction modeling, for each fold, data were split into four training folds and one held-out test fold. LASSO regression was applied only to the training data to select features (i.e., retained features with non-zero coefficients). The learned feature indices were then transferred to the test fold to prevent data leakage, and a support vector regression (SVR) model was trained on the training folds with selected features and evaluated on the test folds. Performance was quantified as the Pearson correlation between predicted and observed values in the test fold. The full CV procedure was repeated 5,000 times using different random partitions to generate bootstrap estimates and was benchmarked against permutation-based null distributions (see Methods for details).

Using the topological measures from all modalities (nine metrics × three modalities), the model significantly predicted task-general LO (predictive $r = 0.38$, $p = 0.0001$) relative to the permutation null (Fig. 4B). The same modeling pipeline also reliably predicted the task-general LR (predictive $r = 0.35$, $p = 0.0004$; Fig. 5A). We further compared the prediction performances of each modality with multimodal combinations (i.e., fMRI+DTI, fMRI+T1, DTI+T1, and all three combined, see Supplementary Fig. S5). We found that across modalities, only DTI topology could produce significant predictions for LO ($p = 0.018$) on its own, while none of the single modalities significantly predicted LR. Notably, although rs-fMRI by itself had the weakest performance and T1 alone cannot generate significant prediction results, in pairwise tests, the rs-fMRI+T1 and DTI+T1 combinations produced significant predictions for both LO (rs-fMRI+T1: $p = 0.049$; DTI+T1: $p = 0.016$) and LR (rs-fMRI+T1: $p = 0.05$; DTI+T1: $p = 0.03$). These results emphasize the complementary strengths of different modalities, with the multimodal data delivering more accurate and consistent predictions than individual modalities.

For LO prediction, network-wise aggregation of LASSO coefficients showed that the DAN and FPN contributed to the prediction above chance (DAN: $p = 0.027$; FPN: $p = 0.042$; Fig. 4C). Prediction models with single-network measures corroborated these effects and also included the limbic network (LIN), with each predicting LO better than chance (see Supplementary Fig. S4A and B; DAN: $p = 0.021$; LIN: $p = 0.026$). For the brain topology metric, nodal local efficiency (NLE) carried the most significant positive coefficients in overall and exceeded the permutation distribution ($p = 0.022$; Fig. 4D). Within the DAN and FPN, different regions showed distinct contributions to the predictions. Within the DAN, the right middle frontal gyrus and parietal regions served as regional hubs (Fig. 4E, upper panel), exhibiting significant predictive weights on the local network metrics (Fig. 4E, lower panel). Predictive contributions from the FPN were mainly localized to the right inferior frontal gyrus, with nodal local efficiency significantly contributing to the predictions (Fig. 4F). Additionally, a reduced predictor set using PCA scores (described in Fig. 3) also showed significant LO prediction (PC1+PC2 prediction model, Supplementary Fig. S4C; $p$ = 0.043), indicating that the balance of cortical segregation (PC1) and subcortical integration (PC2) in network topology underlies individual language learning success, with a prominent contribution from the frontoparietal, attention, and subcortical networks.

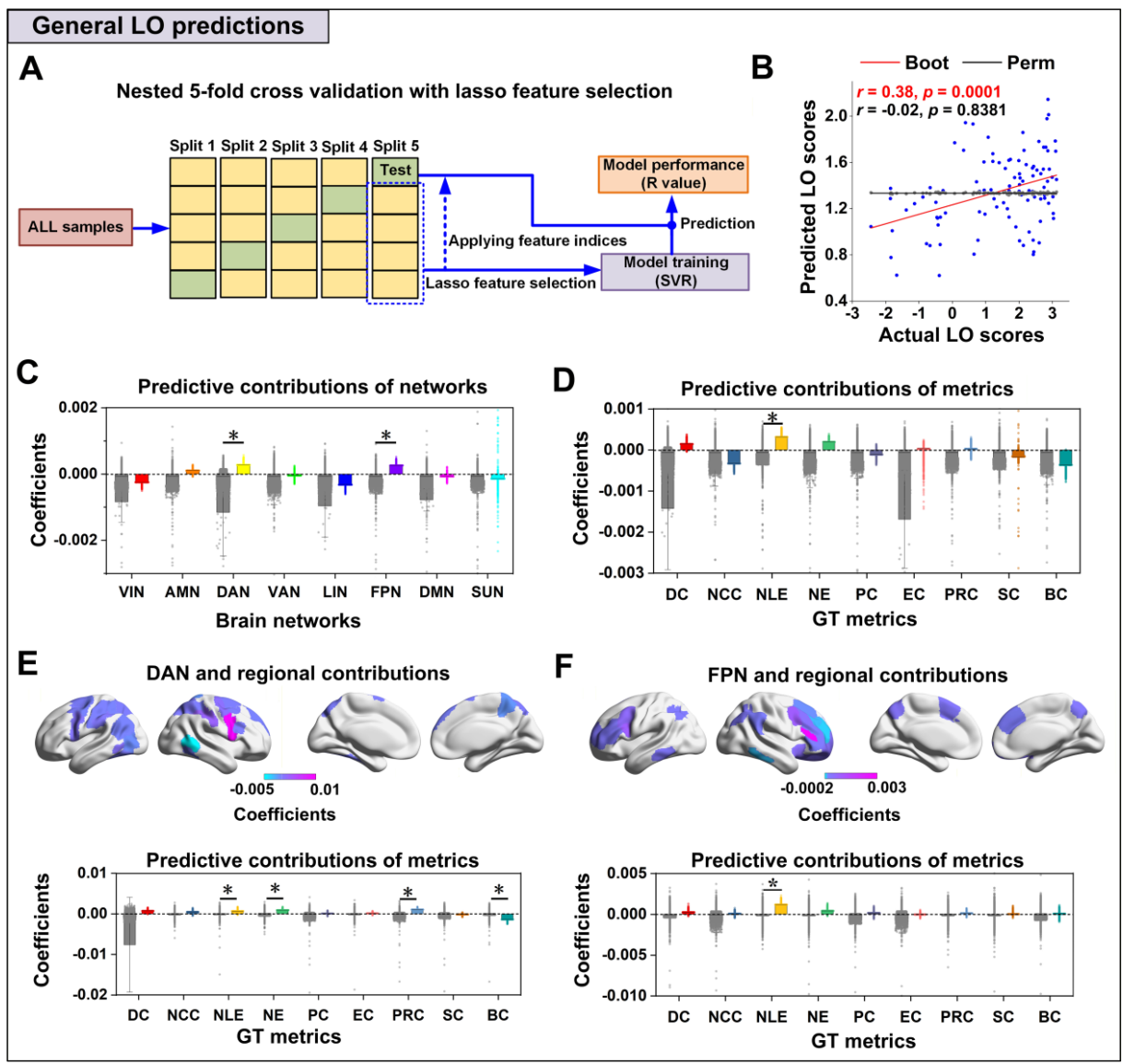

**Fig. 4**. Multimodal topological features predict task-general individual language learning outcomes. **A**. Prediction model construction and validation procedure. A nested 5-fold cross-validation procedure with LASSO feature selection was employed; the selected features from the training sets were then used to construct SVR models, which were subsequently evaluated on the test sets. Performance was the test-fold Pearson correlation ($r$) between predicted and observed values. The full procedure was bootstrapped 5,000 times and compared with permutation nulls. **B**. Learning outcome (LO) prediction modeling using all nine network metrics across three modalities yielded a significant prediction performance (mean $r = 0.38$, $p = 0.0001$ vs. permutation [gray dots]). **C**. Brain network contributions to the LO prediction. By comparing LASSO coefficients from bootstrapping and from permutation, we identified DAN and FPN as significantly above the null (DAN: $p = 0.027$; FPN: $p = 0.043$), contributing to the prediction. Gray bars = permutation-based coefficients. **D**. Overall metric contributions to the LO prediction. Only nodal local efficiency (NLE) exceeds the permutation distribution ($p = 0.022$). **E**. Upper: Surface brain maps display regional contribution weights within the DAN for LO prediction. Lower: metric contributions for DAN. **F**. Upper: Surface brain maps display regional contribution weights within the FPN for LO prediction. Lower: graph-theory (GT) metric contributions for FPN. *, FDR-corrected $p < 0.05$.

For LR prediction, DAN again emerged as a key contributor with coefficients surpassing the null ($p = 0.001$; Fig. 5C). The robust predictive weights were mainly localized to the right middle frontal and parietal regions, identifying them as key hubs (Fig. 5D). Positive coefficients indicate that higher network values lead to faster learning. Metric analyses further revealed regional topological organization, as indicated by the significant contribution from the node cluster coefficient (NCC) and NLE (NCC: $p = 0.017$; NLE: $p = 0.005$). These local topological patterns play important roles, along with the global PageRank centrality (PRC), as significant contributors to the learning speed (PRC: $p = 0.029$; Fig. 5B). Together, these results demonstrate robust prediction of multimodal topological features to predict both final learning success and the speed of learning. Attention- and control-related networks (DAN, FPN), along with their local-connectivity metrics (NLE, NCC) and the global centrality metric (PRC), are especially informative to explain individual differences in language learning.

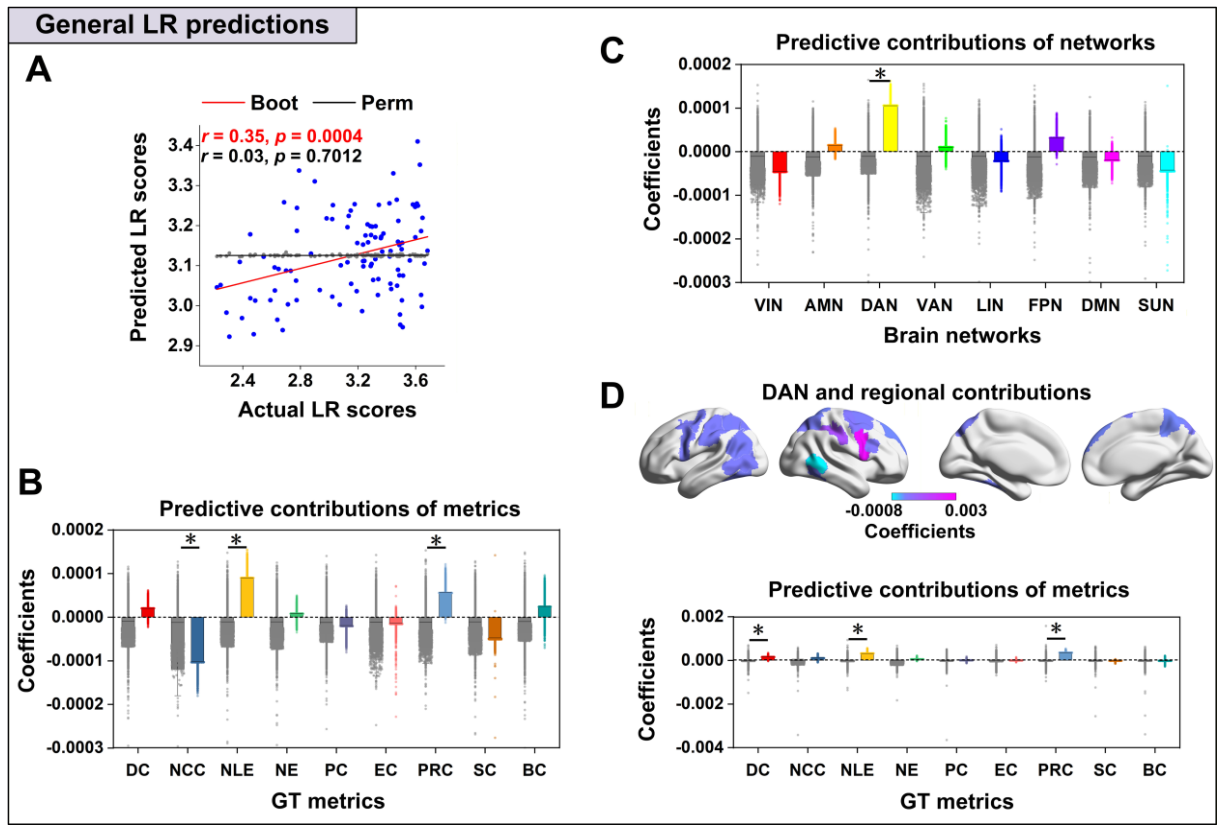

**Fig. 5**. Multimodal topological features predict task-general individual learning rates (LRs). **A**. LR prediction with all network metrics from the three modalities (mean $r = 0.35$, $p = 0.0004$ vs. permutation [gray dots]). **B**. Network metric contributions to LR prediction. NCC, NLE, and PRC surpass permutation-derived coefficients (NCC: $p = 0.017$; NLE: $p = 0.005$; PRC: $p = 0.029$), highlighting roles for local brain connectivity and influence-based centrality. **C**. Network contributions to LR prediction. DAN contributed significantly to the LR prediction ($p = 0.0012$). *, FDR-corrected $p < 0.05$. **D**. Upper: Surface brain maps show regional weights within the DAN for LR prediction. Lower: Metric contributions to the LR prediction within the DAN. Positive coefficients indicate that higher network metric values lead to faster learning.

**Language-component-specific learning predictions**

We also assessed whether multimodal topological features could predict individual LOs and LRs for the five task-specific language components (PC2-PC6, see Fig. S2). These components may correspond to auditory category and word learning (PC2), word learning (PC3), speech vowel category learning (PC4), morphological learning (PC5), and grammar learning (PC6). Predictive performance differed among the components (Fig. 6A), with only LO on the word learning (PC3) being predicted better than chance. The mean predictive performance demonstrated the robustness of LO prediction for the word learning components (mean $r$ = 0.37, $p$ = 0.0001; Fig. 6B).

Examination of model coefficients showed that the frontoparietal network (FPN) and default mode network (DMN) contributed significantly, differing from their permutation distributions (FPN: $p$ = 0.049; DMN: $p$ = 0.007; Fig. 6C). At the metric level, node efficiency (NE) had a negative coefficient, while betweenness centrality (BC) had a positive one; both differed significantly from their null coefficients derived through permutation (NE: $p$ = 0.007; BC: $p$ = 0.016; Fig. 6D). Region-wise weight maps show prediction contributions to FPN and DMN regions. In the FPN, weights were highest in dorsolateral prefrontal and inferior parietal nodes (Fig. 6E). In the DMN, contributive weights were focused in the medial frontal and posterior cingulate/precuneus areas, with additional contributions along the lateral temporal cortex (Fig. 6F). Bar plots (Fig. 5E-F, lower panel) confirmed that DC, NE, and PRC within these networks dominated the metric contributions, but with negative coefficients, indicating that higher network metric values are associated with poorer learning outcomes.

Together, these results suggest that multimodal topology also predicts learner variability in the word learning component, and that this prediction relies on less segregative network organization in the FPN and DMN regions.

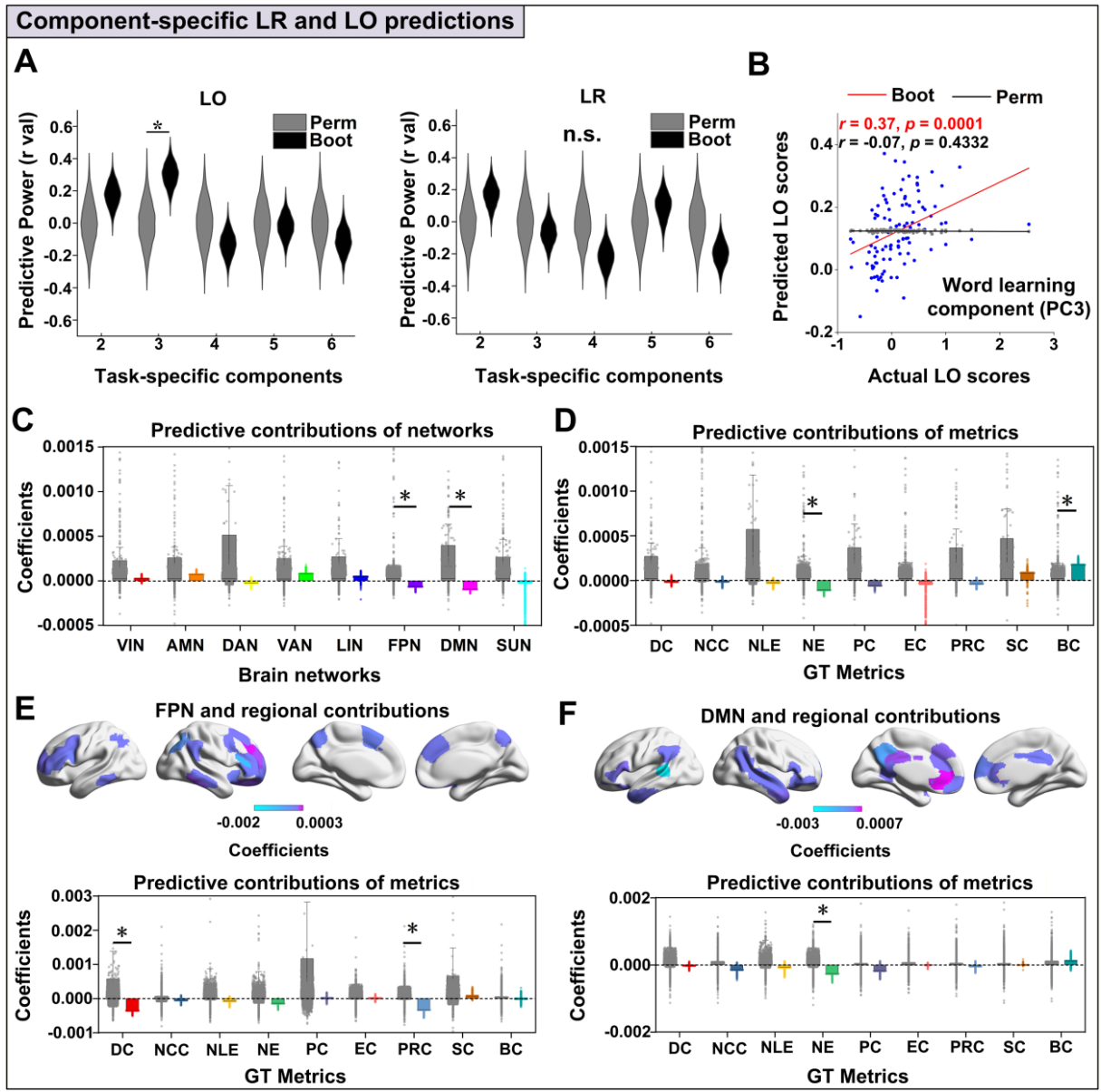

**Fig. 6**. Multimodal topological features predict LO and LR on task-specific components (PC2 to PC6). **A**. Prediction performances on LO and LR of the five task-specific components. Only LO on the word learning component (PC3) significantly predicted better than the permutation chance. **B**. Scatter plots displaying predicted versus observed LO of PC3 (mean $r = 0.37$, $p = 0.0001$), compared with the permutation model (gray dots and line). **C**. Brain network contributions to the LO prediction of the word learning component. FPN and DMN significantly differed from the permutation distribution. **D**. Network metric contributions to LO prediction of the word learning component. BC and NE significantly differed from the permutation distribution. **E**. Upper: Surface brain maps show regional weights within the FPN for the LO prediction. Lower: network metric contributions for the FPN. **F**. Upper: Surface

brain maps show regional weights within the DMN for the LO prediction. Lower: network metric contributions for the DMN. *, FDR-corrected $p < 0.05$.

**Discussion**

We demonstrate that individual differences in adult language learning can be prospectively predicted based on multimodal multi-network brain topology. These findings highlight a component-general language learning ability shared across speech, lexical, and grammatical learning domains. Using seven days of training across six artificial-language learning tasks, we derived a robust task-general component (PC1) from individual learning curves that exceeded permutation-based chance, increased consistently over days, and assigned roughly equal weight to all tasks. This indicates an aptitude-like general language learning capacity rather than task-specific skills (Carroll, 1993; Hedge et al., 2018; Kievit et al., 2017; Spearman, 1904). Multimodal graph-theoretic features computed from functional, structural, and morphological connectivities significantly predicted both final learning outcome (LO) and learning rate (LR), providing systems-level, individualized neuromarkers of language learning that complement previous group-level findings (Feng et al., 2021b; Finn et al., 2015; Rosenberg et al., 2016; Shen et al., 2017). Importantly, distributed cognitive control and attention-related neural networks and their network topological organizations jointly shape how well and how fast adults acquire new linguistic knowledge (Bassett and Sporns, 2017; Duncan, 2010; Fedorenko and Thompson-Schill, 2014).

Group-level network topology reveals segregation-integration network organization between association attention, cognitive control networks, and subcortical regions. Dimension reduction of graph-theory features revealed two dominant dimensions: dimension 1 loaded positively on frontoparietal (FPN), dorsal attention (DAN), and default mode networks (DMN), and negatively on subcortical regions; dimension 2 showed the converse pattern, with strong contributions from the subcortical areas. At the metric level, dimension 1 was dominated by local-connectivity measures (e.g., node local efficiency, degree centrality), whereas dimension 2 emphasized global-connectivity measures (e.g., betweenness centrality, eigenvector centrality). The alignment of DAN, FPN, and DMN suggests a shared topological motif of locally efficient, segregated processing within large-scale association cortex, consistent with their roles in attention, flexible control, and internally guided cognition (Buckner and DiNicola, 2019; Cole et al., 2013; Corbetta and Shulman, 2002). Conversely, the subcortical prominence on global network topology indicates a complementary axis of global communicative influence and hubness that may support rapid gating, reinforcement, and proceduralization processes relevant for sequential structural (e.g., grammar) and category learning (Graybiel and Grafton, 2015). These patterns provide a logical framework (i.e., cortico-subcortical and segregation-integration brain organizations) that connects modality-independent topology to the functional systems involved in language learning.

Critically, predictive modeling uncovered the behavioral relevance of these multi-network topological organizations. For LO, lasso-SVR models integrating all topological features were robustly predictive ($r = 0.38$, $p = 0.0001$), with network-level analyses highlighting the dorsal attention network (DAN), the frontoparietal control network (FPN), and the limbic network as the dominant contributors. For LR, a learning measure that targets learning dynamics (a more difficult learning measure to predict), our prediction models were also significant ($r = 0.35$, $p = 0.0004$), with DAN emerging as the primary network predictor; limbic and subcortical networks also contributed to LR predictions. The dorsal attention networks support sustained vigilance and selective enhancement of task-relevant signals while suppressing distractors (Petersen and Posner, 2012; Vossel et al., 2014). In the context of our language training paradigm, higher DAN weights likely enable learners to: (i) rapidly orient to novel linguistic cues and stimulus-response contingencies, (ii) maintain goal-directed focus across densely packed trials, and (iii) allocate processing resources to the features that carry predictive value (e.g., critical category-defining acoustic modulations). The limbic system and subcortical basal ganglia areas are closely related to memory formation and integrating reward signals (e.g., correct feedback) to consolidate the newly formed linguistic representations (Ashby and O'Brien, 2005; Feng et al., 2019; McClelland et al., 1995; O'Reilly et al., 2014), potentially converting exemplars into abstract categories or rules. These operations may directly regulate the steep, early part of the learning curve when performance is most affected by attentional lapses and thus act as a bottleneck or rate limiter for LR. This interpretation aligns with models of DAN as a mechanism for top-down orienting and priority-map modulation that speeds up evidence gathering (Corbetta and Shulman, 2002; Ptak, 2012).

Additionally, reaching asymptotic accuracy (LO) involves more than just sustained attention. Learners need to flexibly reconfigure strategies, maintain and update task rules, test hypotheses about the underlying structure, and incorporate feedback to solidify correct representations. These functions correspond to the frontoparietal control network (FPN), which has been linked to managing inter-network communication, supporting working-memory gating, and enabling set shifting and model-based exploration. Greater FPN engagement would thus help uncover latent grammatical patterns, choose among competing parses, and solidify the rules into a stable policy involving processes that determine ultimate achievement. This aligns with views of the FPN as a flexible hub that implements control policies and hypothesis testing across domains (Chein and Schneider, 2005; Cole et al., 2013; Duncan, 2010). The FPN may also interact with limbic and subcortical regions involved in hippocampal-medial temporal and basal ganglia circuits responsible for fast encoding, consolidation, and habit formation of newly learned regularities (Davis and Gaskell, 2009; Graybiel and Grafton, 2015). Notably, when using reduced network topology components as predictors, prediction accuracy remained significant for LO, indicating that the brain network segregation-integration between cortical and subcortical regions is a key aspect of brain network organization in learning success; however, it was weaker for LR prediction unless all components were included. This suggests that the temporal dynamics of language learning depend on a broader range of networks and their interactions.

At the metric level, node local efficiency (NLE) emerged as the most reliable predictor of both end-state performance and learning speed, with node cluster coefficient (NCC) and PageRank centrality (PRC) also supporting the prediction of learning. Conceptually, high NLE indicates that a node is embedded in a locally redundant neighbourhood with multiple short alternative routes, enabling fast, robust computation and resilient information relay (Rubinov and Sporns, 2010; Wang et al., 2016). Consistent with prior links between higher local efficiency and superior cognitive performance (Li et al., 2009; Moreira Da Silva et al., 2020), locally efficient topology within dorsal attention and frontoparietal systems likely facilitates rapid evidence accumulation, stable cue weighting, and flexible updating, thereby accelerating early gains while also promoting higher ultimate attainment. Building on this local robustness in network segregation, elevated NCC reflects tightly-knit mesoscale communities that focus on processing, reducing interference, and supporting hypothesis testing, where features that are particularly advantageous during the trial-by-trial adjustments (both grammar and category learning with trial-by-trial feedback) that determine learning rate (Bassett et al., 2015; Betzel and Bassett, 2017). Complementing these segregated computations, higher PRC indexes influential connector roles that broadcast validated local solutions to the broader network, expediting system-wide reconfiguration in the early learning phase (Brin and Page, 1998; Cole et al., 2013). Together, NLE (local robustness), NCC (mesoscale segregation), and PRC (global influence) capture complementary facets of the brain network topology that, in combination, support both rapid adaptation and later consolidation in language learning.

We further found that only the outcomes of the word-learning component among task-specific components were predictable from brain topological patterns. These findings highlight a selective and mechanistically informative link between large-scale network topology and word learning. The significant contributions of DMN and FPN, along with the prominence of betweenness centrality and node efficiency, suggest that hub-mediated routing and global communication play a role in mediating word learning. Intriguingly, we found negative relationships between these networks and their topological measures and word learning outcomes, indicating that learners with less hubness and global communication in the DMN and FPN tend to achieve better outcomes. The fact that single-network models were not predictive emphasizes the importance of cross-network coordination for complex learning, aligning with theories that adaptive behavior results from interactions between internally oriented (DMN) and externally oriented control systems (FPN).

More broadly, our findings support the use of multimodal, topology-aware biomarkers to explain individual differences in language-component-general and -specific learning success and to guide interventions targeting network efficiency and hub integrity, such as cognitive training or neuromodulation. Our findings extend and integrate several strands of literature. First, they offer predictive, individual-level evidence for the involvement of component-general network systems in language learning, complementing region-centric models limited to the frontotemporal language network or areas (Fedorenko and Thompson-Schill, 2014; Hickok and Poeppel, 2007). Second, they demonstrate that multimodal integration enhances sensitivity to person-specific constraints, as structural, functional, and morphological connectomes index partially nonredundant brain topology (Fjell et al., 2015; Forkel et al., 2014; Saygin et al., 2016). Third, they align with network neuroscience, showing that intrinsic topology forecasts behavior in attention, working memory, and learning (Finn et al., 2015; Rosenberg et al., 2016; Shen et al., 2017; Tavor et al., 2016), and they resonate with studies that identify DAN and FPN as scaffolds for adaptive control and transfer across tasks (Cole et al., 2013; Duncan, 2010; Greene et al., 2018). Identifying a general language learning component that can be prospectively predicted from baseline brain topology is especially important for both theory and practice: it indicates a trait-like ability that encompasses phonological, lexical, and grammatical learning and provides a new and task-generalizable target for neuro-informed personalized instruction and rehabilitation (Carroll, 1993; Kievit et al., 2017).

**Limitations and future directions**

Although we predicted a strong overall language learning ability with learners' brain topologies, we could not reliably forecast variations specific to particular task components. Using hierarchical or bifactor models could help distinguish general language aptitude from task-specific learning in explaining individual differences. These models would also enable tests to assess whether language-related frontotemporal brain regions explain residual variation in phonetic, lexical, or morphosyntactic learning beyond what is accounted for by general language ability (Hedge et al., 2018; Kievit et al., 2017).

We used PCA and lasso-based feature selection and validation procedure to identify features while reducing noise effectively; future work could leverage advanced fusion methods (e.g., joint ICA, similarity network fusion, stacked models, or graph neural networks) to capture cross-modal dependencies and nonlinear interactions that may further improve predictive accuracy and interpretability (Scheinost et al., 2019; Sui et al., 2012).

Validation in larger, demographically diverse cohorts, varied language families, and ecologically rich learning contexts is needed. Longitudinal imaging during training can assess whether changes in DAN and FPN local efficiency mediate performance improvements, connecting baseline predictors with neural plasticity mechanisms (Bassett et al., 2015; Gracia-Tabuenca et al., 2024). Rigorous out-of-sample and cross-site validation with motion/confound controls will validate our prediction models and further strengthen translational claims (Scheinost et al., 2019).

**Conclusions**

The baseline multimodal multi-network topology contains predictive information about who will learn languages more effectively and quickly. A general learning component, which captures variance shared across six language learning tasks, was predicted by distributed brain network organizations, with DAN and FPN consistently involved, and nodal local efficiency emerging as a strong cross-modal metric. Group-level topology further revealed that association cortex (DAN/FPN/DMN) aligns with locally efficient organization, while subcortical systems are associated with global influence, reflecting network organization profiles that support individual language learning success. These findings support a multiple learning system theory where attentional and control networks interact with other network regions to determine both the effectiveness and efficiency of language learning, advancing mechanistic understanding and informing personalized, neurobiologically grounded language training approaches.

## Methods and materials

### Participants

One hundred two healthy young adults (72 females, 30 males) aged 18 to 25 years (mean = 21.29, SD = 2.03) were recruited from communities near South China Normal University. One participant was excluded based on quality control criteria for the imaging data. All participants were native Mandarin speakers and were enrolled as university students at the time of the study. None reported a history of hearing or neurological impairments. To minimize the potential effects of prior linguistic experience on learning the artificial language (Brocanto2) (Morgan-Short K, 2007), all participants were screened for previous exposure to Romance languages. Specifically, no subject had formally studied any Romance language, nor had any been immersed in a Romance language environment for more than three weeks. This criterion was applied because the artificial language materials incorporated phonological and structural features typical of Romance languages. The study protocol was approved by the Ethics Committee of the School of Psychology at South China Normal University and the Joint Chinese University of Hong Kong-New Territories East Cluster Clinical Research Ethics Committee. Written informed consent was obtained from each participant before enrollment.

### Artificial language materials

Participants were asked to conduct six artificial language learning tasks, including one auditory category learning, one speech (vowel) category learning, and four Brocanto2 learning tasks over seven days of training. Brocanto2 is an artificial language system designed with a generative grammatical architecture that closely simulates the structural properties of natural language. This design enables the generation, production, and contextual interpretation of novel sentences (Morgan-Short et al., 2012). The lexicon of Brocanto2 comprises 13 lexical items: four nouns (pleck, neep, blom, vode), two adjectives (troise/o, neime/o), one article (li/u), four verbs (kiln, nim, yab, praz), and two adverbs (noyka, zayma). Each lexical item was recorded in isolation. They were subsequently concatenated to form phrases or sentences with a fixed inter-stimulus interval of 300 milliseconds between each word (see Fig. 1A for sample stimuli). Moreover, each sentence in Brocanto2 was visually represented by a corresponding move on a game board displayed on a computer monitor (Fig. 1A). The syntactic and semantic principles of Brocanto2 are illustrated in the example sentence: "neep neime li vode troiso lu yab zayma", which is translated to "The square neep horizontally releases the round vode". This sentence exhibits several core features of the language. Specifically, the four nouns (e.g., vode, neep) are represented as distinct token symbols on the game board. Adjectives such as troiso (round) and neime (square) correspond to circular and square backgrounds, respectively, and occur post-nominally. Articles (lu, li) also follow the noun and exhibit grammatical gender agreement with it. The sentence follows a strict "subject + object + verb" (SOV) order, with the verb yab (release) appearing in sentence-final position. The adverb zayma (horizontally) immediately follows the verb, consistent with the requirement that adverbs appear post-verbally. Each constituent thus corresponds to a defined visual or action-based element within the game context. The morphosyntax training materials consist of 288 phrases or sentences in Brocanto2. Half of these trials (n = 144) were grammatical, while the other half were ungrammatical. More stimulus details could be found in the previous study (Feng et al., 2021b).

**Auditory and speech category training materials**

Participants completed two category-learning tasks: an auditory ripple category-learning task and a speech vowel category-learning task. Both tasks employed an information-integration (II) category structure (Feng et al., 2021a), such that optimal categorization required the integration of two continuous acoustic dimensions rather than reliance on a single, easily verbalizable cue.

*Ripple category-learning stimuli*

Ripple stimuli were generated by applying spectrotemporal modulations to a broadband white-noise carrier. The carrier signal was white noise with energy from 150 Hz up to its fifth octave (4.8 kHz). Each stimulus lasted 500 ms and was sampled digitally at 44.1 kHz. The root-mean-square (RMS) level was set to 80 dB SPL for all ripple stimuli. Each ripple was characterized along two independent acoustic dimensions: spectral modulation frequency and temporal modulation frequency. Spectral modulation frequency, which relates to the density of spectral ripples across the frequency spectrum, ranged from 0.1 to 2 cycles per octave. Temporal modulation frequency, representing the rate of amplitude modulation over time, varied from 4 to 10 Hz. These parameter ranges were chosen because spectrotemporal modulation within this region is strongly represented in the human auditory cortex (Schönwiesner and Zatorre, 2009).

To establish an II category structure, we first defined a two-dimensional abstract perceptual space with both axes normalized from 0 to 1. In this space, four bivariate normal distributions were defined, each representing one category and centered at (0.33, 0.33), (0.33, 0.68), (0.68, 0.33), and (0.68, 0.68). Each distribution had a standard deviation of 0.1 along both axes. From these distributions, we sampled a total of 40 coordinates (10 per category) within the [0, 1] × [0, 1] space. These abstract coordinates were then mapped to physical stimulus parameters as follows: the x-dimension was logarithmically mapped to spectral modulation frequency in the range of 0.1-2 cycles per octave, and the y-dimension was logarithmically mapped to temporal modulation frequency in the range of 4-10 Hz. This process produced 40 ripple stimuli (10 exemplars per category) whose category structure follows an II configuration (Fig. 1A).

*Vowel category-learning stimuli*

Vowel stimuli were designed to replicate the same II category structure as the ripple stimuli, but within a speech space defined by second formant frequency (F2) and vowel duration. Natural recordings of the vowels /iː/ (hereafter /ee/) and /uː/ (hereafter /uu/) served as the endpoint vowels. In addition to these endpoint tokens, intermediate vowels were recorded that perceptually fell between /ee/ and /uu/, produced by gradual changes in tongue position and lip rounding. These endpoint and intermediate recordings served as the basis for all subsequent stimulus manipulations.

The vowel stimuli varied along two primary acoustic dimensions: F2 and duration. Target F2 values ranged from 600 to 2400 Hz, and target durations ranged from 80 to 350 ms. To maintain a relatively consistent overall vowel quality while allowing F2 and duration to carry most of the category-relevant variation, F1 was kept within a narrow range. Specifically, we first sampled 40 values uniformly from the interval [0, 1], then mapped these to F1 frequencies between 380 and 400 Hz. This approach kept F1 relatively steady across tokens while preserving naturalness. All vowel stimuli were normalized to an RMS level of 80 dB SPL.

To construct the II structure for vowels, we reused the same 40 coordinates sampled for the ripple stimuli from the four bivariate normal distributions in the normalized abstract space. For each coordinate, the x-dimension was mapped logarithmically to the F2 range of 600-2400 Hz, and the y-dimension was mapped to the duration range of 80-350 ms (using a monotonic mapping, usually linear for duration). Each coordinate thus specified a target F2 and duration value. For each of these 40 coordinate pairs, we also assigned an F1 value within the constrained range of 380-400 Hz, as described above.

Vowel exemplars were then generated by altering natural recordings through formant shifting and adjusting duration. To ensure consistency and naturalness of the stimuli, we started with an appropriate base token, either an endpoint /ee/ or /uu/ recording, or an intermediate vowel recording. We then produced one exemplar by modifying F2 and duration to match the target values for a specific coordinate. This modified token served as the source for the next exemplar in that category, with F2 and duration adjusted to fit the next target coordinate. By following this process sequentially, we generated four internally consistent "families" of vowels, each representing one category, since all exemplars within a family were derived from a closely related source recording.

For each of the 40 targeted vowel stimuli, F2 was adjusted using formant-shifting techniques to achieve the desired F2 between 600 and 2400 Hz, while preserving the overall vowel identity and timbre as much as possible. Duration was adjusted via time-stretching or compression algorithms designed to minimize pitch distortion and other artifacts, to achieve a duration between 80 and 350 ms. F1 was kept within the narrow 380-400 Hz range to maintain a relatively stable vowel height. Because the stimuli were derived from natural speech recordings, the realized distribution of F2 and duration values could deviate slightly from a perfectly symmetric II layout (e.g., skewing toward one side of the idealized diagram shown in Fig. 1A and Fig. S1), particularly when more extreme parameter combinations would compromise the naturalness of the stimuli.

In both the ripple and vowel tasks, the four categories were defined as clusters in a two-dimensional stimulus space: spectral modulation frequency × temporal modulation frequency for ripple sounds, and F2 × duration for vowel sounds. The optimal decision boundaries in these spaces (see Fig. 1A and Fig. S1) require listeners to integrate information along both dimensions, consistent with an information-integration category structure.

**Training procedure**

Participants completed six language-learning tasks over seven consecutive days (Fig. 1B). Each daily session included: (i) an auditory category-learning (ACat) task, (ii) a speech category-learning (SCat) task, (iii) a Brocanto2 word-learning (Word) task followed by a word test, and (iv) three Brocanto2 morphosyntax-learning tasks focused on morphological rules (MR), phrase structure (PS), and sentence structure (SS). On Day 1, all tasks except SCat were performed during fMRI scanning; from Days 2 to 7, training was conducted outside the scanner. On the final day, participants also completed a generalization test for each of the three morphosyntax tasks. The Brocanto2 training protocol has been detailed elsewhere (Feng et al., 2021a, 2021b). The current training battery was designed to assess learning across multiple language components, including basic auditory categories, speech categories, lexical items, and morphosyntactic structures.

*Auditory and Speech Category Training*

In both the ACat and SCat tasks, participants learned to categorize sounds into four categories. On Day 1, auditory stimuli were delivered through MRI-compatible headphones, while visual feedback was projected onto a screen and viewed via a mirror mounted on the head coil. On subsequent days, the same behavioral setup was used outside the scanner, with stimuli presented through standard laboratory headphones and visual feedback displayed on a computer monitor. Task structure and timing remained consistent across days.

Each trial was time-locked to a sparse-sampling fMRI acquisition sequence on Day 1. A 1700 ms image acquisition was followed by an 800 ms silent gap, during which a single auditory stimulus (ripple or vowel, depending on the task; see above) was presented. Participants were instructed to categorize the stimulus into one of four categories using a four-button handheld response box. They were encouraged to respond as accurately as possible within the trial window. After the response, trial-wise feedback was provided to facilitate category learning. Corrective feedback (e.g., "Correct: this is Category 1" or "Incorrect: this is Category 3") appeared on the screen for 750 ms. The feedback onset was jittered relative to sound onset, with a variable interval of 2-4 seconds between stimulus presentation and feedback, to decorrelate stimulus- and feedback-evoked BOLD responses. Each trial lasted 7.5 seconds (equivalent to 3 TRs) and included a jittered post-feedback interval of 0.45-5 seconds. Additionally, null trials with only central fixation (no sound and no response) were interspersed throughout each run to improve the measurement of single-trial activation.

Within each daily ACat and SCat session, trials were organized into six training blocks. The order of stimuli within each block was randomized for each participant. Before beginning the experimental blocks, participants completed a brief practice session to familiarize themselves with the button-to-category mappings and response procedure. Apart from the presence or absence of MRI scanning and the type of headphones used, the task structure, timing, and feedback schedule were identical across the seven training days.

*Brocanto2 Training*

Brocanto2 is an artificial miniature language that has been extensively used to study language learning and processing (Feng et al., 2021a, 2021b). Each daily Brocanto2 session began with a word-learning task (Word) and was followed by three morphosyntax training tasks targeting morphological rules (MR), phrase structure (PS), and sentence structure (SS). On Day 1, all Brocanto2 tasks were performed during fMRI scanning; on subsequent days, the same procedures were administered outside the scanner.

During the word-learning phase, participants were exposed to spoken Brocanto2 lexical items paired with their corresponding visual referents. Each trial presented an auditory word together with a visual symbol (or configuration of symbols) that depicted its meaning (e.g., game pieces or moves in a simplified "chess-like" environment). Participants were instructed to learn the associations between the novel words and their referents. The word-learning phase was followed by a brief test phase, in which participants were required to select the correct referent for a heard word, allowing for the assessment of their emerging lexical knowledge.

Morphosyntax training was implemented using grammaticality judgment tasks (GJTs) that targeted three types of grammatical knowledge (i.e., MR, PS, and SS). In each GJT, participants listened to Brocanto2 phrases or sentences while viewing corresponding visual "moves" or configurations on a chessboard-like display. On each trial, the auditory stimulus and the visual display were presented concurrently, followed by a prompt instructing participants to judge whether the utterance was grammatically acceptable in the language. Participants indicated "grammatical" or "ungrammatical" using button presses. After each judgment, they received trial-by-trial corrective feedback indicating whether their response was correct, thereby allowing them to implicitly acquire the morphosyntactic rules governing the language.

For each grammar type, the GJT consisted of two experimental blocks per day, resulting in a total of six blocks. Each block contained a randomized mixture of 24 grammatical and 24 ungrammatical stimuli (48 trials per block). For each trial, the duration of stimulus presentation depended on the length and complexity of the phrase or sentence. Within each grammar type, the grammatical and ungrammatical items were matched in length and acoustic properties as closely as possible. After stimulus onset, a response window was provided during which participants made their grammaticality judgment. Corrective feedback followed each response. To disambiguate neural responses to the stimulus and to the feedback in the fMRI data, the interval between the participant's response and the onset of feedback, as well as the inter-trial interval, was jittered across trials.

The Brocanto2 procedures (word learning and morphosyntax GJTs) were identical on Day 1 (inside the scanner) and on Days 2-7 (outside the scanner) in terms of stimuli, trial structure, feedback, and block organization; only the testing environment and recording of brain activity differed. On the final training day, participants completed an additional generalization test for the MR, PS, and SS grammar types, in which novel grammatical and ungrammatical sentences were used to assess their ability to apply learned rules to previously unseen items.

**Multimodal imaging data acquisition**

MRI data were collected using a Siemens 3T Tim Trio system with a 20-channel head coil at the Brain Imaging Center of South China Normal University (Fig.1C). High-resolution T1-weighted structural images were acquired with an MPRAGE sequence: 176 slices, TR = 1900 ms, TE = 2.53 ms, flip angle = 9°, and voxel size = 1×1×1 mm³. Resting-state functional MRI (rs-fMRI) was performed using a multi-band EPI sequence: TR = 2000 ms, TE = 30 ms, 58 slices, slice thickness = 2 mm, flip angle = 90°, and a matrix size of 112×112. Each participant was scanned for approximately 5.5 minutes, capturing 150 volumes. Diffusion tensor imaging (DTI) data were collected with an echo-planar imaging sequence: 64 slices, TR = 9800 ms, TE = 89 ms, slice thickness = 2 mm, flip angle = 90°, and a matrix size of 128×128. The DTI protocol included 65 diffusion-gradient orientations (b = 700 s/mm²) plus one initial b0 image.

**Analysis of artificial-language learning performances**

**Principal component analysis**

As mentioned earlier, the learning scores in our study may reflect both a component-general language learning trajectory and component-specific patterns. These components may be supported by different neural network systems. To distinguish between general learning trajectories and component-specific patterns in the behavioral data, we applied principal component analysis (PCA) to the seven-day learning scores. Specifically, the three-dimensional dataset (101 participants × 7 days × 6 tasks) was transformed into a two-dimensional matrix of size 707 × 6. PCA was then performed across the six tasks to identify orthogonal components that represent shared variance across tasks and task-specific components. To further evaluate the significance of these principal components (PCs), we conducted a permutation test with 10,000 iterations. During each iteration, the data were randomly permuted, and PCA was recalculated. This process generated a null distribution of the explained variance for each component. Additionally, to assess the stability and reliability of the identified component structure, we performed a bootstrap analysis with 10,000 resamples. In each bootstrap iteration, participants were randomly sampled with replacement from the original dataset, and PCA was reapplied to the resampled data. The median explained variance from the bootstrap distribution for each component was then compared to the 95th percentile of the null distribution generated by the permutation test. A component was considered statistically significant if its explained variance was greater than the null distribution in 95 percent of cases.

**Learning outcome and learning rate estimation**

In this study, we examined two key metrics to characterize individual learning trajectories derived from the principal components of performance scores across six tasks: learning outcome (LO) and learning rate (LR). Both metrics were based on PC scores obtained from the PCA applied to the six-task behavioral data collected over the training period. Learning outcome was defined as each participant's performance on the final day (day 7) of training. We used the generalization test score for the three morphosyntax learning tasks and the mean accuracy on day 7 for the other three tasks. Learning rate was estimated for each participant by fitting a linear-log function to their principal component scores across seven training days. We also added a chance level baseline before day 1 to better reveal the learning gains from chance to the ultimate performance. Specifically, a regression model was fitted where the natural logarithm of the day number served as the independent variable $x$ and the PC scores as the dependent variable $y$, as shown in Equation 1.

$$y = a \log(x) + b \qquad (1)$$

The slope $a$ derived from the fitted model was used as an index of the learning rate, reflecting the steady increase from chance performance to the final outcomes on day 7. Together, LO and LR provide complementary insights into the learning dynamics and ultimate learning success.

**Multimodal brain connectome construction**

**Morphological connectome based on sMRI**

For each participant, T1-weighted structural MRI images were co-registered to Montreal Neurological Institute (MNI) space using Advanced Normalization Tools (ANTs) and resampled to an isotropic resolution of 1 mm³. A total of 47 radiomic features were subsequently computed for each of the 246 regions defined by the Brainnetome atlas (Fan et al., 2016). These regions were grouped into 8 major brain networks, including the visual network (VIN), auditory-motor network (AMN), dorsal attention network (DAN), ventral attention network (VAN), limbic network (LIN), frontal-parietal network (FPN), default mode network (DMN) and subcortical network (SUN). Prior to connectome construction, feature values were normalized across brain regions within each individual using min-max scaling (Zhao et al., 2021). To reduce redundancy, features with high correlation ($r > 0.9$) to any other feature were removed, resulting in 25 measures and a feature matrix of 246×25 per subject (Zhao et al., 2021). The Pearson correlation coefficients were calculated using the radiomics feature vector from each pair of regions of interest (ROIs). Finally, a 246×246 morphological connectome matrix was obtained for each subject.

**Functional connectome based on rs-fMRI**

The standard fMRI preprocessing steps were carried out using the DPABI toolbox (https://rfmri.org/DPABI). Specifically, the first 10 volumes were discarded, followed by slice-timing correction. Next, each volume of a participant was realigned to the mean image using a linear transformation. After realignment, the mean functional image was co-registered to the corresponding T1 image, which had been segmented into gray matter, white matter, and cerebrospinal fluid with a unified segmentation method (http://www.fil.ion.ucl.ac.uk/spm). Finally, the functional images were resampled to 3 mm × 3 mm × 3 mm and normalized into MNI space using DARTEL (Ashburner, 2007). To reduce the effect of noise, the images were smoothed with a 4-mm FWHM Gaussian kernel and bandpass filtered between 0.01 and 0.1 Hz. The nuisance regression was used to remove irrelevant variable interferences, including the Friston-24 parameters, white matter signal, and cerebrospinal fluid signal.

**Structural connectome based on DTI**

The PANDA toolbox (Pipeline for Analyzing Brain Diffusion Images, Beijing Normal University, http://www.nitrc.org/projects/panda/) was used for DTI data processing. First, the raw image was resampled and cropped to enhance computational efficiency. Then, head motion and eddy current-induced distortion were corrected to improve image quality. After that, DTI metrics such as fractional anisotropy (FA) and mean diffusivity (MD) were estimated. To compute the structural connectivity matrix, deterministic fiber tracking was performed using the fiber assignment by continuous tracking (FACT) algorithm. Tracking was terminated if the turning angle exceeded 45° or the FA value dropped below 0.2. The Brainnetome atlas was used to parcellate the brain into distinct regions. A structural connectivity matrix was then constructed, where the average FA value of streamlines connecting each pair of regions defined the weight of the corresponding edge.

**Graph-theory metrics**

To measure the brain's topological properties at the node level, we further performed graph theory analysis based on the connectome matrices from the T1, DTI, and rs-fMRI data using the GRETNA toolbox (http://www.nitrc.org/projects/gretna). Specifically, multiple sparsity thresholds (5%-40%, interval=1%) were used to obtain a series of networks for each subject (Chen et al., 2019). To delineate the comprehensive network properties, nine types of node-wise graph-theory metrics were calculated for each region: (I) Degree Centrality (DC); (II) Node Cluster Coefficient (NCC); (III) Node Local Efficiency (NLE); (IV) Node Efficiency (NE); (V) Participant Coefficient (PC); (VI) Eigenvector Centrality (EC); (VII) PageRank Centrality (PRC); (VIII) Subgraph Centrality (SC); (IX) Betweenness Centrality (BC). Finally, the area under the curve (AUC) was calculated for each metric across various sparsity levels and used for further analysis. AUC has been shown to be highly sensitive to topological changes in the brain and is independent of threshold selection (Wang et al., 2009; Zhang et al., 2011).

To better integrate these various features from multiple neuroimaging modalities and reveal the structure of these network metrics, we used PCA to extract the primary component from the graph-theory metrics. Specifically, after calculating the graph-theory metrics, a total of 27 feature measures were obtained (9 network metrics × 3 modalities). PCA was performed on the data matrix with dimensions 74,538 (i.e., 246 regions × 101 subjects × 3 modalities) by 9 metrics, resulting in 9 principal components (PCs). To identify statistically significant components, a permutation test and a bootstrapping with 10,000 iterations were conducted. In each iteration, the features were randomly shuffled across participants, and PCA was rerun to create a null distribution of explained variance for each PC. Only those PCs whose explained variance exceeded the corresponding null distribution were considered significant. These significant PCs were also used as combined multimodal features in subsequent analyses.

**Predictive modeling analysis**

To assess the predictive power of the multimodal features, we used a support vector regression (SVR) model with a nested five-fold cross-validation framework. The least absolute shrinkage and selection operator (Lasso) method was applied to select the most informative features for predicting learning profiles. The entire process was randomly repeated 5000 times to ensure the robustness of the results. We calculated Pearson's correlation between predicted and actual scores to evaluate prediction performance.

Additionally, a permutation test with 5000 iterations was performed. In each permutation, the relationship between input features and prediction labels was randomly shuffled, breaking the true associations in the data. This generated a null distribution of performance metrics, assuming no real relationship exists. The actual model performance, based on 5000 bootstrap iterations within the 5-fold cross-validation, was compared to this null distribution. Specifically, the median Pearson correlation coefficient from the bootstrapped cross-validation was evaluated against the permutation-based null distribution. The empirical p-value was computed as the proportion of permutation results that exceeded the actual median performance. This rigorous, non-parametric statistical approach confirms whether the observed model performance is significantly better than chance, thus enhancing the validity and reliability of our results.

# Supplementary Materials

**Supplementary Figures**

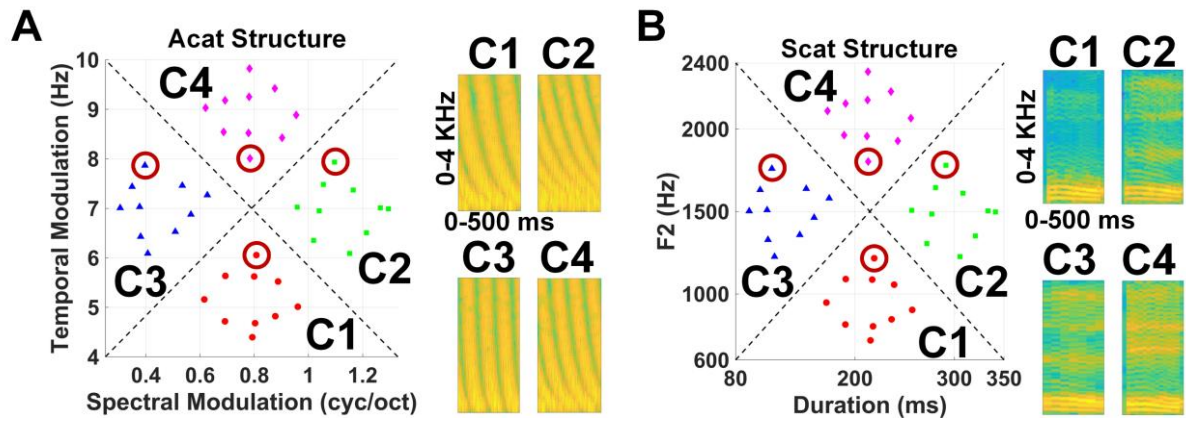

**Fig. S1**. Category structure of four categories and spectrograms of sample sounds used in the ripple auditory (A) and speech vowel (B) category learning experiments. **A**. Category structure, sound distribution, and sample spectrograms in the ripple auditory category learning experiment. **B**. Category structure, sound distribution, and sample spectrograms in the speech vowel category learning experiment. Each dot in the 2D acoustic space represents a sound. The spectrograms of the representative samples highlighted by red circles are displayed.

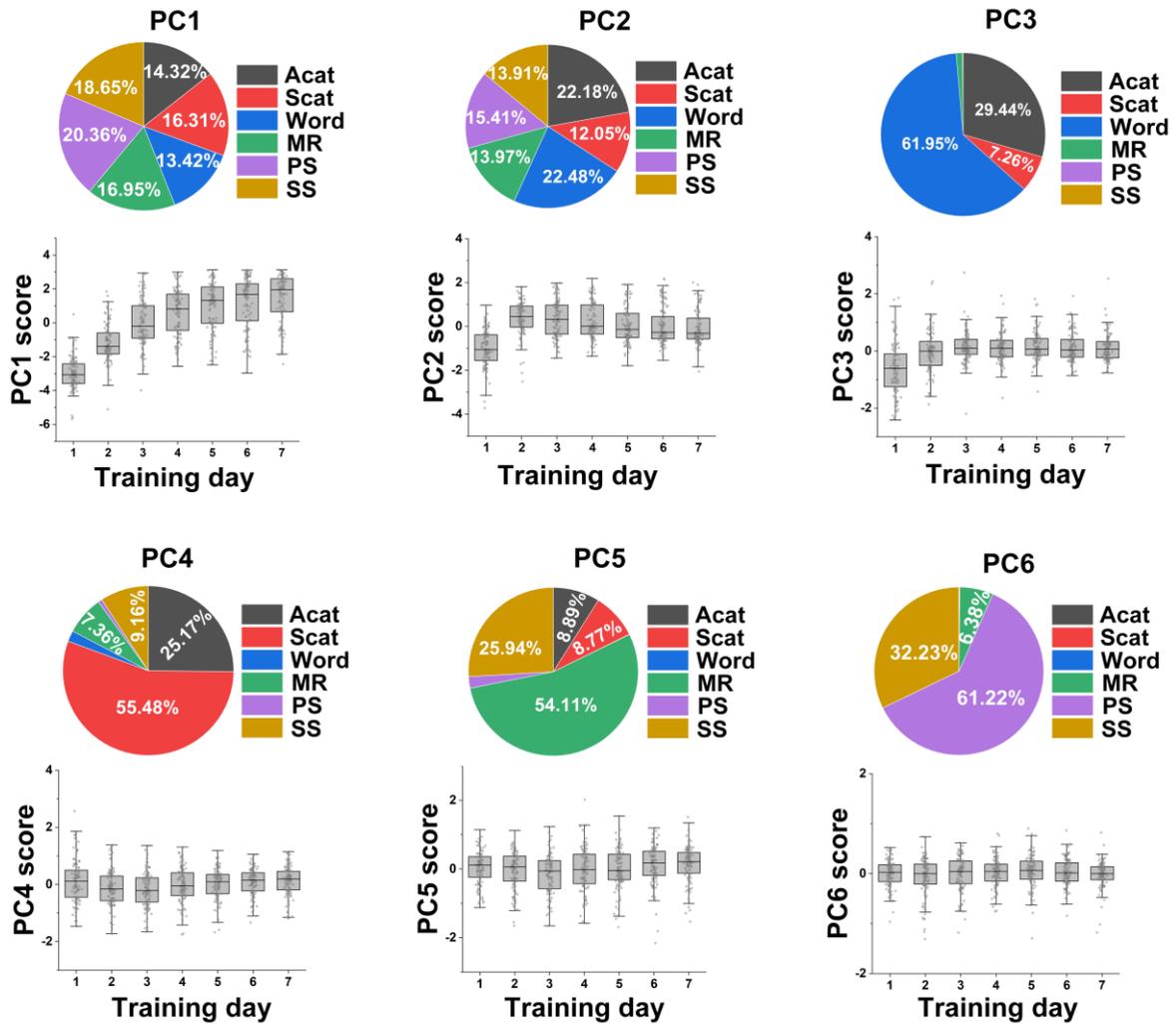

**Fig. S2**. Language learning components PC1 to PC6 and their PC scores (contributions) are shown across days using boxplots and across tasks with a pie chart. PC1 represents a general learning pattern; PC2 represents category and word learning; PC3 represents word learning; PC4 represents vowel speech category learning; PC5 represents morphological learning; PC6 represents grammar learning.

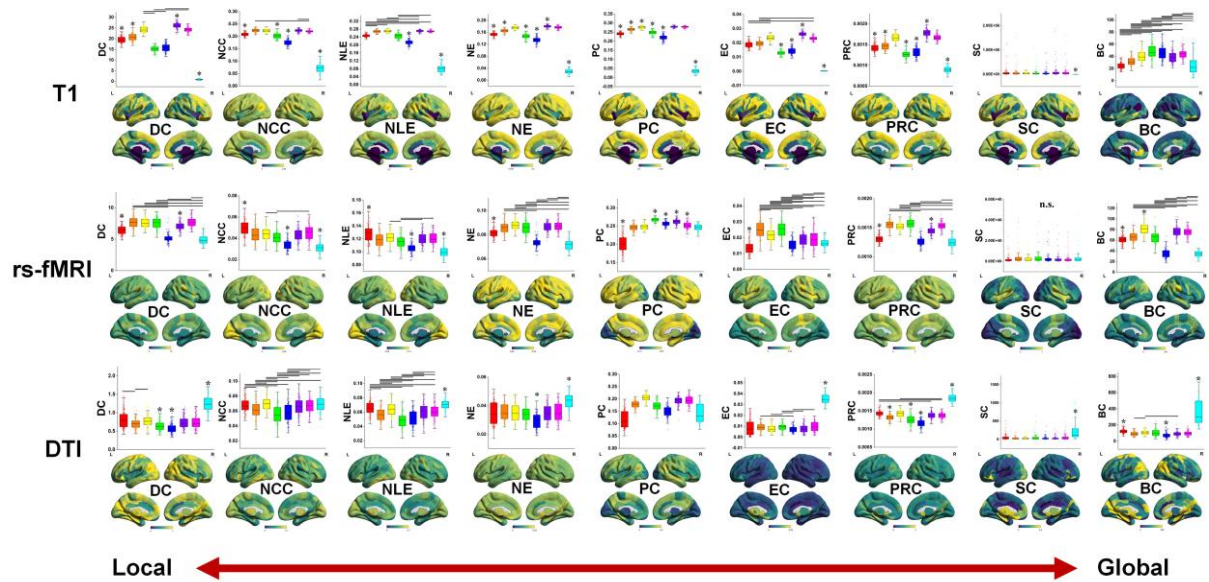

**Figure S3.** Comparative topography of nine graph-theoretic network metrics across morphological, functional and structural connectomes. Regional (bottom) and network-level (top) distributions for each metric are shown. Metrics: Degree Centrality (DC), Node Cluster Coefficient (NCC), Node Local Efficiency (NLE), Node Efficiency (NE), Participant Coefficient (PC), Eigenvector Centrality (EC), PageRank Centrality (PRC), Subgraph Centrality (SC), and Betweenness Centrality (BC). Asterisks denote networks that differ significantly from all others; horizontal bars indicate significant pairwise differences between the connected networks (FDR $p < 0.05$). L/R indicates hemispheres.

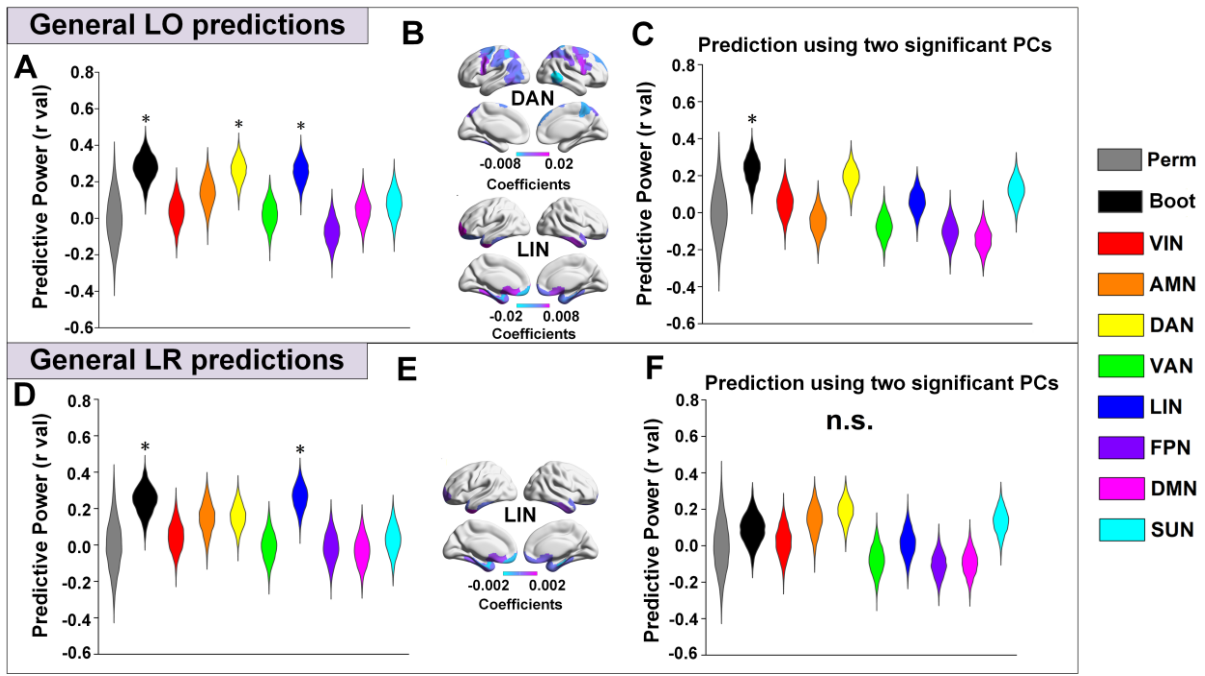

**Fig. S4**. Predictive performance of individual brain networks for predicting task-general language learning outcomes (LO) and learning rate (LR). **A**. LO prediction performance for each network using all the multimodal topological features. **B**. Surface maps show regional weights within the DAN and LIN regions for LO prediction. **C**. LO prediction for each network using only the two significant principal components of the topological features. **D**. LR prediction performance for each network using all the multimodal topological features. **E**. Surface maps show regional weights within the LIN regions for LR prediction. **F**. LR prediction for each network using the two principal components (PC1-PC2). Perm = permutation-based prediction performance; Boot = bootstrapping-based prediction performance. Network abbreviations: VIN = visual network; AMN = auditory-motor network; DAN = dorsal attention network; VAN = ventral attention network; LIN = limbic network; FPN = frontal-parietal network; DMN = default mode network; SUN = subcortical network.

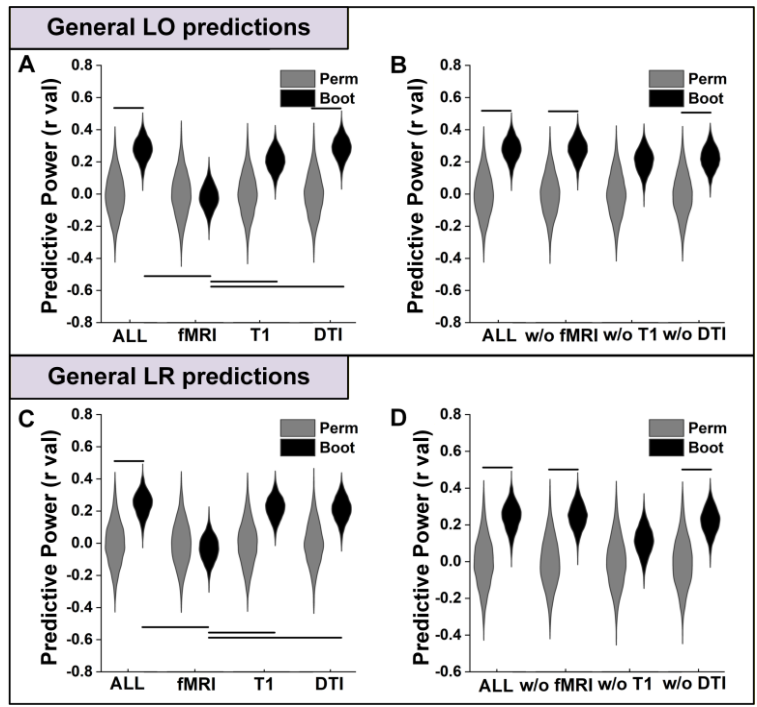

**Fig. S5**. Predictive performance of various combinations of multimodal topological features for predicting task-general (PC1) language learning outcomes (LO) and learning rate (LR). **A**. Predictive performance for LO using features from each modality. **B**. Predictive performance for LO using features from two modalities. w/o = without (i.e., remove one modality). **C**. Predictive performance for LR using features from each modality. **D**. Predictive performance for LR using features from two modalities. Perm = permutation-based prediction performance; Boot = bootstrapping-based prediction performance. Horizontal bars indicate significant pairwise differences between the connected predictive results.

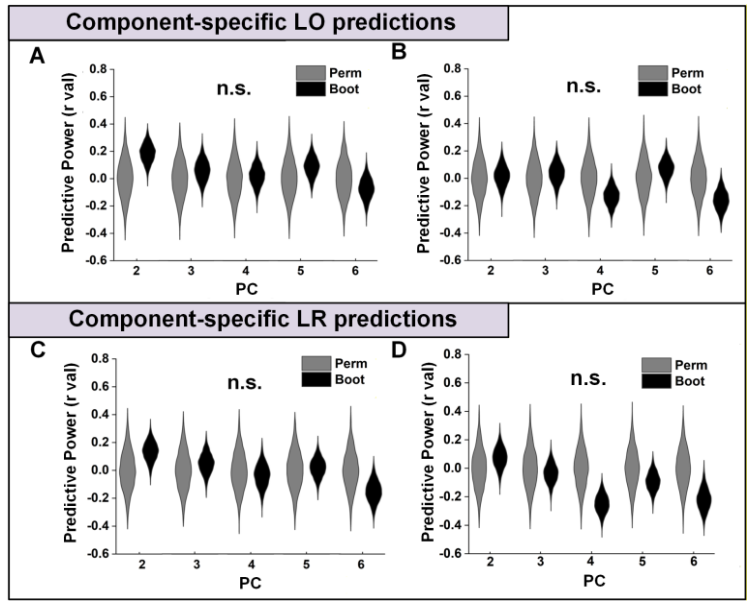

**Fig. S6**. Predictive performance of task-specific language components PC2 to PC6 using multimodal topological features. **A.** Predictive performance for LO of PC2-PC6 using all nine principal components. **B.** Predictive performance for LO of PC2-PC6 using PC1 and PC2 derived from multimodal features. **C.** Predictive performance for LR of PC2-PC6 using all nine principal components. **D.** Predictive performance for LR of PC2-PC6 using PC1 and PC2 derived from multimodal features. Perm = permutation-based prediction performance; Boot = bootstrapping-based prediction performance.